\documentclass{elsart}

\usepackage{graphicx}

\usepackage{amssymb}
\newtheorem{theorem}{Theorem}

\begin{document}

\begin{frontmatter}



\title{Quantum information cannot be split into complementary parts}


\author[GAT,ITP]{D.L. Zhou\corauthref{cor}},
\corauth[cor]{Corresponding author.} \ead{dz34@mail.gatech.edu}
\author[MIT]{B. Zeng},
\ead{zengbei@MIT.EDU}
\author[GAT,ITP]{L. You}
\ead{ly14@mail.gatech.edu}

\address[GAT]{School of
Physics, Georgia Institute of Technology, Atlanta, GA 30332, USA}

\address[ITP]{Institute of Theoretical Physics, The Chinese Academy
of Sciences, Beijing, 100080, China}

\address[MIT]{Department of Physics, Massachusetts Institute of
Technology, Cambridge, MA 02139, USA}

\begin{abstract}
We prove a new impossibility for quantum information (the
no-splitting theorem): an unknown quantum bit (qubit) cannot be
split into two complementary qubits. This impossibility, together
with the no-cloning theorem, demonstrates that an unknown qubit
state is a single entity, which cannot be cloned or split. This
sheds new light on quantum computation and quantum information.
\end{abstract}

\begin{keyword}
quantum entanglement\sep quantum information
\PACS 03.67.-a\sep 03.65.-w\sep 89.70.+c
\end{keyword}

\end{frontmatter}

\section{Introduction}

As a century old theory, quantum mechanics has provided the most
effective description of the physical world. Recently, new
discoveries were found for its applications to information and
computation science \cite{Nielsen}, \textit{e.g.}, the efficient
prime factorization of larger numbers \cite{Shor} and the perfectly
secure quantum cryptography \cite{Bb84}. These, and related
developments, have highlighted a general theme that quantum
mechanics often makes impossible tasks in the classical world
possible. Conversely, some possible operations in the classical
world become impossible in the quantum world \cite{Pati2}. For
example, an unknown quantum state cannot be perfectly cloned
\cite{Wootters,Dieks}, while copies of an unknown quantum state
cannot be deleted except for being swapped into the subspace of an
ancilla \cite{Pati}.

The principle of linear superposition of states is an important
feature of quantum mechanics. A significant consequence is that an
unknown quantum state cannot be perfectly cloned, which has been
known for quite some time \cite{Wootters,Dieks}. This impossibility
can also be understood from the causality requirement that no signal
can be transmitted faster than the speed of light, even with the aid
of nonlocal quantum resource such as entanglement. With the rapid
development of quantum information science in recent years, we have
come to realize the essential role of this simple, yet profound,
limitation in quantum information processing, especially in quantum
cryptography \cite{Bb84}. Intuitively, the no-cloning theorem
implies there exists an essential difference between one copy and an
ensemble of such copies of an unknown quantum state. One cannot
obtain any information from only one copy of the quantum state
without any prior knowledge of the state. Extensive research has
focused on the no-cloning theorem related topics in quantum
information science \cite{Yuen,Barnum,Duan}. Recently, Pati
discovered another important theorem of impossibilities for an
unknown quantum state based on the principle of linear
superposition: no linear transformations on two copies of an unknown
quantum state can delete a copy except for being swapped into an
ancilla state \cite{Pati}.

In this letter, we show that yet another theorem of impossibilities
exists:  quantum information of an unknown qubit cannot be split
into two complementing qubits, \textit{i.e.}, the information in one
qubit is an inseparable entity. Our paper is organized as follows:
in Sec. II we present our no-splitting problem in terms of a common
scenario from quantum secret sharing. We show that if our discussion
is restricted to only product pure final states, then the
no-splitting statement is apparently valid. Following, in Sec. III,
we consider the nontrivial case of the no-splitting problem,
\textit{i.e.}, for pure entangled final states. We then present a
no-splitting theorem for a two-qubit case and argue that the
no-splitting theorem also should be true in more general cases.
Finally, we discuss several effects and applications of our
no-splitting problem and point out possible future directions.

We note that Pati and Sanders have independently developed a similar
idea -- the no-partial erasure of quantum information -- in a recent
paper \cite{Pati3}. They claim that our non-splitting theorem
becomes a straightforward corollary of their no-partial eraser
theorem.  This, however, is not the case. As demonstrated in their
example of Eq.(8), if the final state is allowed to be a mixed state
(for example due to entanglement with an ancilla), their no-partial
eraser becomes invalid. On the contrary, the final pure state can
contain entanglement between of the two (complementary) qubits for
our theorem, thus our result must supersedes their no-partial
erasure theorem.  In fact, as we show in Sec. II, the no-partial
erasure theorem is valid for product states, but not for the more
general case of entangled states in Sec. III.  We emphasize that the
possible existence of entanglement between the two qubits is what
makes our theorem on non-splitting of quantum information more
important.

\section{The No-splitting problem}

We start by presenting our non-splitting idea in terms of a common
scenario from quantum secret sharing: we assume that Alice and Bob
want to store and share a secret, say, an unknown spatial direction
of a qubit on the Bloch sphere, specified by its Euler angle
$(\theta,\phi)$. If this secret is initially held by Alice, she can
simply send the unknown value of $\theta$ or $\phi$ to Bob in the
classical world, and this would accomplish one simple scheme of the
secret sharing as they now each possess the complementary part of
the secret $\theta$ or $\phi$. However, this scheme as well as all
other classically allowed more sophisticated schemes is impossible
in the quantum world.

With the pseudo-spin representation on the Bloch sphere, the
unknown qubit initially held by Alice can be denoted as
\begin{equation}
|v(\theta, \phi)\rangle_A=\cos\frac {\theta} {2} |0\rangle_A +
\sin \frac {\theta} {2} e^{i \phi}|1\rangle_A.\label{unkstat}
\end{equation}
In terms of this state, the no-cloning theorem says that there
exists NO unitary transformation $\mathcal{U}$ such that
\begin{equation}
\mathcal{U}|v(\theta, \phi)\rangle_A|w\rangle_B =|v(\theta,
\phi)\rangle_A|v(\theta, \phi)\rangle_B,
\end{equation}
where $|w\rangle_B$ denotes an arbitrary given state of the
ancilla qubit $B$. The no-deleting theorem of Pati states that
there exists NO unitary transformation $\mathcal{U}$ either to
achieve the following
\begin{equation}
\mathcal{U}|v(\theta, \phi)\rangle_A|v(\theta,
\phi)\rangle_B|w\rangle_C =|v(\theta,
\phi)\rangle_A|x\rangle_B|y\rangle_C,
\end{equation}
where for clarity we have assumed two copies of the unknown state.
And, $|x\rangle_B$ and $|y\rangle_C$ are any known states.

A restricted form of the no-splitting theorem, \textit{the two real
parameters $(\theta,\phi)$ contains in one qubit can not be split
into two complementary qubits in a product state}, can be
mathematically stated as follows. There does not exist any unitary
transformation $\mathcal{U}$ such that
\begin{equation}
|\Psi\rangle_{AB}:=\mathcal{U}|v(\theta,\phi)\rangle_A|w\rangle_B=|x(\theta)\rangle_A
|y(\phi)\rangle_B.\label{distribu}
\end{equation}
When we use the linearity of $\mathcal{U}$ (from quantum mechanics),
the plausible forms for states on the right hand side of Eq.
(\ref{distribu}) are
\begin{eqnarray}
|x(\theta)\rangle_A&=&\cos \frac {\theta} {2} |x_1\rangle_A+ \sin
\frac {\theta} {2} |x_2\rangle_A,\\
|y(\phi)\rangle_B&=&|y_1\rangle_B+e^{i\phi}|y_2\rangle_B,
\end{eqnarray}
with un-normalized states $|x_1\rangle_A$, $|x_2\rangle_A$,
$|y_1\rangle_B$, and $|y_2\rangle_B$, all independent of $\theta$
and $\phi$. It is an easy exercise to conclude this kind of linear
transformation cannot exist in quantum mechanics by comparing the
LHS with the RHS of Eq. (\ref{distribu}).

The above version of no-splitting theorem for product pure final
states is valid also for more general cases with higher dimensions
and more parameters. This restricted version can indeed be derived
from the no-partial erasure theorem (Theorem 4) in Ref. [11], but
the converse is not true (Corollary 5 in Ref. [11]). We will show in
the following section that the no-partial erasure theorem is invalid
for the more general case of entangled pure final states. In
contrast, our no-splitting theorem remains valid for both cases.

\section{No-splitting theorem}

The above restricted version of the theorem is limited to
separable pure states in the RHS of Eq. (\ref{distribu}). More
generally, $|\Psi\rangle_{AB}$ can take the form of an entangled
pure state. For example, when the unitary transformation
$\mathcal{U}$ corresponds to a control-NOT gate with qubit $A$ as
the control qubit and $|w\rangle_B=|0\rangle_B$, we obtain
\begin{eqnarray}
|\Psi\rangle_{AB}&=&\frac {1} {2} \left(\cos \frac {\theta} {2}
|0\rangle_A+ \sin \frac {\theta} {2}
|1\rangle_A\right)\left(|0\rangle_B+e^{i\phi}|1\rangle_B\right) \nonumber\\
&+& \frac {1} {2} \left(\cos \frac {\theta} {2} |0\rangle_A- \sin
\frac {\theta} {2}
|1\rangle_A\right)\left(|0\rangle_B-e^{i\phi}|1\rangle_B\right),\nonumber \\
\label{examp}
\end{eqnarray}
which consists of coherent superpositions where each contains a
split state of $\theta$ and $\phi$. Does this example point to a
failure of our non-splitting idea when $|\Psi\rangle_{AB}$ is an
entangled state? No. In fact, in this case we only need to examine
the reduced density matrix of qubit $A$ and $B$, respectively. For
the state (\ref{examp}), the reduced density matrix for qubit $A$
({or} $B$) is
\begin{equation}
\rho_{A(B)}=\cos^2{\frac {\theta}
{2}}|0\rangle_{A(B)}\mbox{}_{A(B)}\!\langle0|+\sin^2{\frac
{\theta} {2}}|1\rangle_{A(B)}\mbox{}_{A(B)}\!\langle 1|,
\end{equation}
both independent of $\phi$. Thus, the above example does not provide
a counterexample to our non-splitting idea.

It is also straightforward to show that the no-partial erasure
theorem of Pati and Sanders \cite{Pati3} is no longer valid in this
case, since for the state (\ref{examp}), simply discarding one qubit
will result in a mixed state with parameter $\theta$. This
observation is trivial because a simple measurement in the
computational basis will erase the information of $\phi$. On the
other hand, as shown by the above observation, our no-splitting
theorem remains valid. We formulated our no-splitting idea into the
following theorem, which constitutes the central result of this
letter.
\begin{theorem}
There exists no two-qubit unitary transformation $\mathcal{U}$
capable of splitting an unknown qubit. In mathematical terms, the
transformed state is
\begin{equation}
|\Psi\rangle_{AB}:=\mathcal{U}|v(\theta,\phi)\rangle_A|w\rangle_B,
\label{transstat}
\end{equation}
where $|v(\theta,\phi)\rangle_A$ is defined in Eq. (\ref{unkstat}),
and $|w\rangle_B$ is an arbitrarily given pure state of qubit $B$.
This theorem then states that
\begin{eqnarray}
\textrm{
tr}_B\left(|\Psi\rangle_{AB}\mbox{}_{AB}\!\langle\Psi|\right)&=&\rho_A(\theta)\label{requir1}
\end{eqnarray}
and \begin{eqnarray}
 \textrm{
tr}_A\left(|\Psi\rangle_{AB}\mbox{}_{AB}\!\langle\Psi|\right)&=&\rho_B(\phi)\label{requir2}
\end{eqnarray}
cannot be satisfied simultaneously.
\end{theorem}
We now prove this general result.\\
\textbf{Proof}: Inserting Eq. (\ref{unkstat}) into Eq.
(\ref{transstat}), we obtain
\begin{eqnarray}
|\Psi\rangle_{AB}=\cos \frac {\theta } {2}
\mathcal{U}|0\rangle_A|w\rangle_B+\sin \frac {\theta} {2}
e^{i\phi} \mathcal{U}|1\rangle_A|w\rangle_B.
\end{eqnarray}
Applying the Schmidt decomposition of a two-qubit pure state, we
immediately find
\begin{equation}
\mathcal{U}|1\rangle_A|w\rangle_B=r_0|\tilde{0}\tilde{0}\rangle_{AB}
+r_1|\tilde{1}\tilde{1}\rangle_{AB},
\end{equation}
where $|\tilde{0}\rangle_{A(B)}$ and $|\tilde{1}\rangle_{A(B)}$ are
the corresponding orthogonal basis states of the Schmidt
decomposition for qubits $A$ and $(B)$,  and $r_0$ and $r_1$ are
real parameters which satisfy the normalization condition
\begin{equation}
r_0^2+r_1^2=1.
\end{equation}
Because the state $\mathcal{U}|0\rangle_A|w\rangle_B$ is
orthogonal to state $\mathcal{U}|1\rangle_A|w\rangle_B$, we deduce
that
\begin{equation}
\mathcal{U}|0\rangle_A|w\rangle_B=\alpha r_1
|\tilde{0}\tilde{0}\rangle_{AB} -\alpha r_0
|\tilde{1}\tilde{1}\rangle_{AB}
+c|\tilde{0}\tilde{1}\rangle_{AB}+d|\tilde{1}\tilde{0}\rangle_{AB},
\end{equation}
where  $\alpha$, $c$, and $d$ are generally complex. They satisfy
the normalization condition
\begin{eqnarray}
|\alpha|^2+|c|^2+|d|^2=1.
\end{eqnarray}
The conditions of Eqs. (\ref{requir1}) and (\ref{requir2}) are
summarized in the following equivalent set of equations:
\begin{eqnarray}
d^* r_0 &=&0,\\
c r_1&=&0,\\
c^* r_0&=&0,\\
d r_1&=&0,\\
\alpha r_0 r_1&=&0,\\
|\alpha|^2 r_1^2 +|d|^2-r_0^2&=&0,\\
c^* \alpha r_1 -d \alpha^* r_0&=&0.
\end{eqnarray}
Suppose $r_0\neq 0$, then $c=d=\alpha r_1=0$, but $r_0^2=|\alpha|^2
r_1^2 +|d|^2=0$; therefore, $r_0=0$, which is contradictory. Now
assume $r_0=0$, which leads to $r_1\neq 0$ and $c=d=0$, then
$|\alpha|^2=( {r_0^2-|d|^2})/ {r_1^2}=0$, thus
$|\alpha|^2+|c|^2+|d|^2=0$. Again this is contradictory. Thus, there
is no self-consistent solution to Eqs. (\ref{requir1}) and
(\ref{requir2}), \textit{i.e.}, we have completed the proof of our
theorem.

When $|\Psi\rangle_{AB}$ is a product pure state, Eqs.
(\ref{requir1}) and (\ref{requir2}) reduces to Eq. (\ref{distribu}).
Theorem $1$ further indicates that the information of the amplitude
($\theta$) and the phase ($\phi$) cannot be split into two qubits by
any two-qubit unitary transformation, even for more general
(entangled) pure final two qubit states.

We speculate that the no-splitting theorem is valid for more general
cases of higher dimensional Hilbert spaces with more parameters.
This is based on the observation that the number of constraining
equations grows faster than the number of parameters; hence, in
general no solution could be expected just as we show above for the
case of two qubits.

\section{Applications and future directions}

It has been debated that some tasks of quantum information
processing can only be implemented in real Hilbert space or
restricted to equatorial states (states with the same amplitude on
all the computational basis but different phases). However, the
tasks never would work in the complete complex Hilbert space, for
example, Pati's remote state preparation protocol \cite{pati} and
its higher dimensional generalizations \cite{zz}, the $(2,2)$
quantum secret sharing protocol with pure states \cite{cleve}, and
Yao's self-testing quantum apparatus \cite {yao}. Our theorem,
therefore, provides a stronger evidence that all such tasks can
never be implemented in the whole complex Hilbert space, even
including the potential effort of transferring complex states into
real or equatorial ones. Furthermore, Grover's algorithm
\cite{grover} only calls for rotations of real angles, and Shor's
algorithm \cite{shor} requires discrete Fourier transform which only
needs transformation between equatorial states. Our theorem thus
implies that in some cases, the restricted quantum information and
computation schemes in real or equatorial space may have the same
power \cite{Ber}, or even more power, than schemes in the whole
complex Hilbert space.

Interestingly, despite such strong restrictions from the restricted
version of our no-splitting theorem or the no-partial erasure
theorem \cite{Pati3} that there exists even no probabilistic
approach for splitting or partially erasing an unknown state, the
converse procedure, \textit{i.e.}, to combine two states
\begin{equation}
\cos{\frac{\theta}{2}}|0\rangle+\sin{\frac{\theta}{2}}|1\rangle,
\frac{1}{\sqrt{2}}(|0\rangle+e^{i\varphi}|1\rangle)
\end{equation}
into one can be easily accomplished. As a simple example, we give
the following protocol starting from
\begin{equation}
\left(\cos{\frac{\theta}{2}}|0\rangle+\sin{\frac{\theta}{2}}|1\rangle\right)\otimes
\frac{1}{\sqrt{2}}(|0\rangle+e^{i\varphi}|1\rangle),
\end{equation}
executing a parity detection measurement ($ZZ$), followed by an XOR
gate, then discarding the ancillary qubit, we will reach either
\begin{equation}
\cos{\frac{\theta}{2}}|0\rangle+\sin{\frac{\theta}{2}}e^{i\varphi}|1\rangle,
\end{equation}
or
\begin{equation}
\cos{\frac{\theta}{2}}e^{i\varphi}|0\rangle+\sin{\frac{\theta}{2}}|1\rangle,
\end{equation}
both with the probability of $1/2$. We believe this interesting
observation will shed light on future investigations of the
``quantum nature" of quantum information.

\begin{figure}[htbp]
\begin{center}
\includegraphics[width=3.25in]{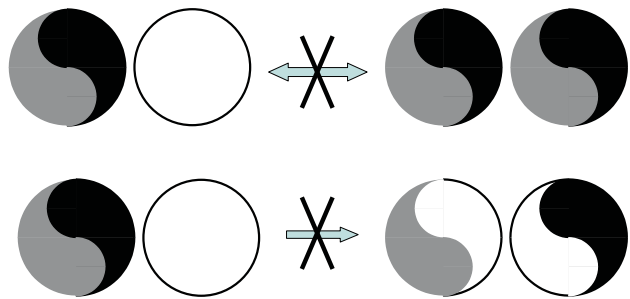}
\end{center}
\caption{Schematic illustration of our result for no-splitting is
in the second row, as compared to the no-cloning and its inverse
no-deleting theorems in the first row. The unknown initial qubit
is represented by the ying-yang circle together with the known
ancilla qubit represented by the empty circle on the left.}
\label{fig1}
\end{figure}

In summary, we have shown that the unknown information of one copy
of a qubit cannot be split into two complementary qubits, whether
the final pure state of the two qubits is separable or entangled.
Our result demonstrates the inseparable property for quantum
information in terms of an unknown single qubit and is schematically
illustrated in Fig. \ref{fig1}. Together with the no-cloning
theorem, the no-splitting theorem shows that one qubit is an entity
that corresponds to the basic unit in quantum computation and
quantum information.

We thank Mr. P. Zhang, Ms. J. S. Tang, Prof. C. P. Sun, and Prof.
Z. Xu for useful discussions. This work was supported by the US
National Science Foundation and by the National Science Foundation
of China.

\end{document}